\documentstyle[aps,prl,twocolumn,epsfig,graphics]{revtex}

\begin{document}

\title{Dissipative spin-boson model and Kondo effect in low 
dimensional quantum gases}
\author{A. Recati,$^{1,2)}$ P.O. Fedichev,$^{1)}$ W. Zwerger$^{3)}$, 
J. von Delft$^{3)}$, and P.
Zoller$^{1)}$}

\address{$^{1}$ Institute for Theoretical Physics, University of Innsbruck,
A--6020 Innsbruck, Austria\\
$^2$ Dipartimento di Fisica, Universit\`a di Trento and INFM, I-38050 Povo, 
Italy\\
$^{3}$ Sektion Physik, Universit\"at M\"unchen, Theresienstr. 37/III, D-80333 M\"unchen,
Germany.}

\maketitle

\begin{abstract}
We show that ultracold atoms confined in a quasi-1d
trap can be used to study the dissipative spin-boson model with {\it tunable} 
tunneling splitting and strength of dissipation. 
It is shown that with realistic parameters this system allows to study 
the crossover from coherent Rabi dynamics to incoherent tunneling.
\end{abstract}

\vspace{0.5cm}

An effective two state system (spin 1/2) coupled to a bath of harmonic 
oscillators 
is one of the most important models to describe the effect of 
dissipation in 
quantum mechanics \cite{spinboson:leggett,weiss}. The associated 
spin-Boson 
Hamiltonian was extensively studied for a wide range of problems, most 
notably the issue of decoherence for superpositions of macroscopically 
distinct states \cite{spinboson:leggett} or various condensed matter 
realizations like the 
tunneling of interstitials in solids which are coupled to phonons or to 
electron-hole excitations in a metal \cite{hydrogen}. In spite of its 
apparent 
simplicity, the spin-Boson model exhibits a very rich behavior, which 
crucially depends on the low frequency weight of the effective 
environment 
spectrum. It ranges from a simple damped oscillation described by the 
phenomenological Bloch equations well known e.g. from NMR to 
complete localization in the so called ohmic dissipation case \cite{spinboson:leggett}. 
The latter is in fact closely related to a generalized 
Kondo-Hamiltonian, 
a paradigmatic model for quantum impurity problems. 

In most of these applications, the precise strength and frequency 
spectrum 
of the environment depends on microscopic details and cannot be changed 
externally. This is the case e.g. in the context of macroscopic quantum 
coherence, where the final observation of coherent superpositions of 
counter-propagating currents in a SQUID \cite{squid1,squid2} requires 
extremely low dissipation whose origin is still not 
completely 
understood on a microscopic level.  Similarly, charged impurities in a 
metal 
are subject to ohmic dissipation through electron-hole excitations,   
however 
their strength is typically very small, depending on the precise 
scattering phase shifts at the Fermi energy. 

Taking advantage of recent experimental progress in cooling and trapping of low dimensional
quantum gases \cite{salomon2002,Ertmer}, we show that cold atoms can be used 
to realize an ohmic spin-Boson model with a {\it tunable} strength of both the 
interaction and the effective tunneling amplitude. 
This  allows to study the complete range of dynamics from coherent 
oscillations to incoherent dynamics in the Kondo regime and 
finally to localization. The oscillator bath in this case arises 
from the coupling to the low energy excitations of a Bose condensate, 
which produces an ohmic spectrum in the one-dimensional case, 
where gapless quantum liquids exhibit universal low energy properties.

The spin-boson model describes a two-level system (a spin-$1/2$,
characterized by the Pauli matrix operators $\sigma $), interacting
with a bath of harmonic oscillators (phonons) according to the following
Hamiltonian:
\begin{equation}
H=-\frac{\Delta}{2} \sigma _{x}+\sum _{q}\omega _{q}b_{q}^{\dagger }b_{q}+
\frac{\sigma_z}{2}\sum _{q}\lambda _{q}(b_{q}+b_{q}^{\dagger }),
\label{eq:Hspinboson}
\end{equation}
where $\Delta $ is the {}``tunneling'' amplitude (the Rabi frequency
in the case of a {}``free'' two-level system), $b_{q}$ and $b_{q}^{\dagger }$
are the annihilation and creation operators of the phonon modes characterized
by a momentum $q$ and the dispersion relation $\omega _{q}=uq$,
where $u$ is the velocity of sound. Formally eliminating the oscillator
part of the Hamiltonian, it can be shown that the reduced dynamics of the spin
is completely determined by the effective density of states 
 \[
J(\omega )=\sum _{q}\lambda _{q}^{2}\delta 
(\omega -\omega _{q})\sim \omega ^{s}\]
for sufficiently small values of $\omega $. Depending on the value
of the exponent $s$ there are several different regimes. The
most interesting situations arises in  the so called ohmic case $s=1$. The latter is
is equivalent to $1/r^2$-Ising model, which is known to have a phase transition 
even in 1d \cite{ising1d}.

\begin{figure}[]
\centerline{\epsfig{file=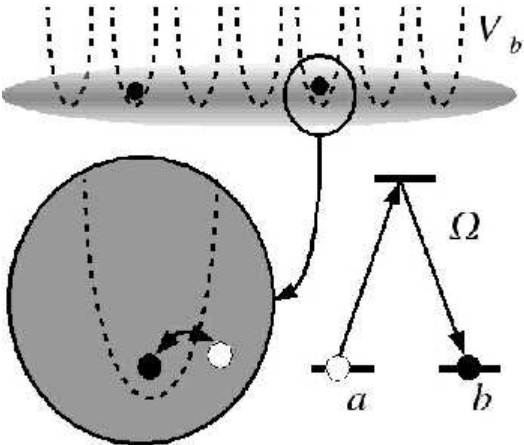,width=7cm}}
\caption{The suggested experimental setup. The Bose-liquid of atoms ``a''
(white circles) is confined in a shallow quasi-1d trap (gray). The atoms
``b'' (black circles) are localized by a separate trap (dashed line).
The transition between the atoms is performed by a laser light.}
\label{cap:setup}
\end{figure}

The spin-boson Hamiltonian (\ref{eq:Hspinboson}) can be implemented
in a setup involving a sample of trapped cold atoms. Consider the
two distinct hyperfine states (let us call them the state $a$ and
$b$ respectively) of the same atom confined within a quasi-1d trap
(see the Fig. \ref{cap:setup}). The atoms of the two bosonic species
are trapped by the two different external potentials, $V_{a}(x)$
and $V_{b}(x)$, respectively. Both potentials are assumed to have
a high frequency of the radial motion, so that the motion of the particles
can be considered as quasi-1d (this implies that both the temperature
$T$ and the characteristic interparticle interaction do not exceed
the characteristic frequency of the transverse confinement, $\omega _{\perp }$).
Most of the atoms are in the state $a$ and form a dense quantum liquid
(the number of atoms in the state $a$ is large $N_{a}\gg 1$). The
potential $V_{b}$ is assumed to have the form of an optical lattice
with very well separated and tightly confining potential wells. Both
potentials should spatially overlap to allow the interaction between
the atomic species. The optical lattice should have fairly low filling:
 $N_{b}\ll N_{a}$,
so that the wavepackets of atoms $b$ do not overlap with each other
and hence have no cooperative interactions with the quantum liquid.
At the same time, $N_{b}$ should be sufficiently large, to facilitate
(ideally non-destructive) measurement.

The interatomic interaction is described by the three interaction
constants $g_{\beta \beta ^{\prime }}$, with $\beta, \beta' = a,b$ 
characterizing the interaction between the atoms $a$, $b$ and between the 
atoms $a$ and $b$. All the interactions are assumed to be short-range.
The values of coupling constants are then related to the 3d scattering
amplitude in a fairly complicated way and may contain resonances.
In the simplest case, when the scattering length, corresponding to
the scattering of the components $\beta $ and $\beta ^{\prime }$,
$a_{\beta \beta ^{\prime }}$, is much smaller than the radial ground
state size $l_{\perp }\sim \omega _{\perp }^{-1/2}$ (hereafter we
use units such that $\hbar =m=1,$ where $m$ is the mass of an atom),
the effective 1d {}``scattering length'' is given by a simple relation:
$g_{\beta \beta ^{\prime }}=2\pi a_{\beta \beta ^{\prime }}/l_{\perp }^{2}$. 
The value and even
the sign of the interaction constants $g_{\beta \beta ^{\prime }}$
may also be changed by applying external magnetic field (Feshbach
resonances, see \cite{Schreck} and references therein), or simply by 
changing the trap aspect ratio
(i.e. by varying $\omega _{\perp }$, see \cite{petrov:1Dinteraction}).

The on-site repulsion between the atoms $b$ is $U_{bb}\sim g_{bb}/l_{b}$,
where $l_{b}$ is the characteristic size of the ground state wavefunction
$\psi _{b}(x)$ localized within a single potential well of potential
$V_{b}$. The value of $U_{bb}$ should be large compared with the
characteristic energy of interaction within and with the Bose-liquid
of atoms $a$: $g_{aa}\rho ,|g_{ab}\rho |\ll U_{bb}$ (but still $U_{bb}\alt \omega _{\perp }$).
Then, only one atom of the type $b$ can be trapped on the same lattice
site at the same time. Accordingly, any state of the atoms $b$ can
be modeled by the corresponding state of a spin-$1/2$ system, namely:
the spin-up state corresponding to, say, the state with a single atom
$b$, whereas the spin-down state corresponds to the situation with
no atoms $b$ on a certain site. 

To model the exchange interaction we suggest to put the atomic system
into the far-detuned laser light, providing a transition between the
internal states $a$ and $b$. The amplitude of the Rabi flip $\Omega $
can, in principle, be spatially inhomogeneous. The part of the Hamiltonian
describing the interaction of atoms $b$ with the Bose-liquid can
be presented in the form of the following spin Hamiltonian: 
\begin{equation}
H_{b}=\frac{g_{ab}}{2}(1+\sigma _{z})\int dx\psi _{b}^{2}\rho _{a}+\int dx\Omega (\Psi _{a}\psi _{b}\sigma _{+}+h.c.),
\label{eq:bHam}
\end{equation}
 where $\hat\Psi_a(x)$ is the annihilation operator valued-field of an atom $a$
at the point $x$, $\rho_a=\hat\Psi_a^\dagger\hat\Psi_a$ is the liquid density operator.
We note, that in order to ensure a single occupancy of the lattice
site the interaction between the atoms $b$ should be repulsive, $g_{bb}>0$
and sufficiently large. At the same time, the interaction of the atoms
$a$ and $b$ can have arbitrary sign (provided that the number of
atoms $b$ is sufficiently small). The first term of (\ref{eq:bHam})
contains a time-independent phonon displacement operator (the identity
part of $1+\sigma_{z}$ operator), which can be always integrated
out from the system's action and only shifts the effective Hamiltonian
by an additive constant. 

Now we turn to the description of the atoms $a$. To implement the
Hamiltonian (\ref{eq:Hspinboson}) we need to exploit a situation
when the dispersion relation for the low energy excitations of the
Bose liquid of atoms $a$ is linear in momentum, i.e. has no gap.
In the long wavelength limit a Bose-gas is a Luttinger Liquid (LL), and 
is described by
the Haldane hydrodynamics Hamiltonian \cite{haldane:prllut}:
\begin{equation}
H=\frac{1}{2\pi }\int dx(v_{J}\partial \phi ^{2}+
v_{N}(\partial \theta -\pi \rho_a)^{2}),\label{eq:LLHam}
\end{equation}
 where $\rho_a$ is the equilibrium liquid density, $v_{J}=\pi \rho_a$ and
$v_{N}=\kappa /\pi \rho_a^{2}$ where $\kappa $ is the compressibility of the liquid.
The fields $\phi $ and $ \theta $ describe the superfluid velocity
and the density fluctuations, respectively.

The LL model (\ref{eq:LLHam}) can be diagonalized by the following
transformation: 
\begin{equation}
\phi (x)=-i\sum _{q}\left|\frac{2\pi }{qLK}\right|^{1/2}e^{iqx}(b_{q}-b_{-q}^{\dagger })
\label{eq:phidef}
\end{equation}
\begin{equation}
\theta (x)=-i\sum _{q}\left|\frac{2\pi K}{qL}\right|^{1/2}sign(q)e^{iqx}(b_{q}+
b_{-q}^{\dagger }),
\label{eq:thetadef}
\end{equation}
 where $L$ is the sample length. Accordingly, the Hamiltonian (\ref{eq:LLHam})
takes the form:
\begin{equation}
H_{LL}=v_{s}\sum _{q}|q|b_{q}^{\dagger }b_{q},
\label{eq:HLL0}
\end{equation}
 where: $v_{s}=(v_{N}v_{J})^{1/2}=(\kappa /\rho_a)^{1/2}$. As
expected, the spectrum of the lowest energy excitations is characterized
by a single sound velocity $u=v_{s}$ (cf. Eq.(\ref{eq:Hspinboson})).

The actual values of the phenomenological parameters $v_{s}$ and
$K=(v_{J}/v_{N})^{1/2}$ in the LL Hamiltonian (\ref{eq:LLHam})  
depend on $\rho_a$, the interaction
$g_{aa}$, and on the parameters of the external potential $V_{a}$.
The requirement of the absence of a gap in the excitation spectrum sets a 
number of restrictions. First of all, the interparticle interaction has to be
repulsive ($g_{aa}>0$). Second, the trapping potential should not
allow the formation of the so-called Mott-insulator state. The Bose-field
operator can be expressed as 
\begin{equation}
\hat\Psi_{a}\sim \left(\rho_a+\frac{\partial \theta }{\pi }\right)^{1/2}
\exp (i\phi )\sum _{m=even}\exp (im\theta ),
\label{eq:psidef}
\end{equation}
where the dimensionless proportionality coefficient depends on non-universal
short interparticle separation properties of the Bose-liquid. 
On the contrary, the expression for the density in terms of
the field $\Pi $ does not have this ambiguity: 
\begin{equation}
\rho _{a}(x)=(\rho_a+
\Pi (x))\sum _{m=even}\exp (i2m\theta (x)).
\label{eq:rhodef}
\end{equation}
In the  case of a trapped Bose-gas, the Luttinger parameter
$K$ depends on the interaction in the combination $\gamma =g_{aa}/\rho_a$
\cite{haldane:prllut}.
In particular, for weak interactions ($\gamma \rightarrow 0$) 
$K(\gamma )\approx \pi /\gamma ^{1/2}$
and thus can be very large. In the other limit, i.e. when the interaction
is strong, $\gamma \rightarrow \infty $, $K(\gamma )\approx 1$ (Tonks gas limit, 
\cite{1dteo}).

Eqs.(\ref{eq:psidef}) and (\ref{eq:rhodef}) for the Bose-field operator
and the particle density can be used to rewrite the interaction Hamiltonian
(\ref{eq:bHam}). The role of the terms with various values of $m$
is quite different. Since the phase $\theta $ contains a quickly
oscillating term $\pi \rho_ax$, all the contribution containing
$\exp (im\theta )$ with $m\neq 0$ average out under the integral
sign in Eq.(\ref{eq:bHam}), provided that the localization length
$l_{b}$ of the wavefunction $\psi _{b}$ is sufficiently large: $\pi \rho_al_{b}\gg 1$.
Then, in the limit $q\rightarrow 0$ we can put $qx\approx 0$ everywhere
in Eqs.(\ref{eq:phidef}) and (\ref{eq:thetadef}). Finally, keeping
only the terms with $m=0$ in Eqs.(\ref{eq:psidef}) and (\ref{eq:rhodef})
we can derive the following representation of the Hamiltonian (\ref{eq:bHam}):
\[
H_{b}=\frac{g_{ab}}{2\pi }\sigma _{z}\sum _{q}
\left|\frac{2\pi K}{L}q\right|^{1/2}(b_{q}+b_{q}^{\dagger })+\]
\begin{equation}
+\tilde{\Omega }\rho_a^{1/2}l_{b}^{1/2}(\sigma _{+}
\exp (i\phi(0))+h.c.),\label{eq:Hbnew}\end{equation}
 where \[
i\phi(0)=\sum _{q}\left|\frac{2\pi }{qLK}\right|^{1/2}(b_{q}-b_{q}^{\dagger }),\]
 and $\tilde{\Omega }\sim \Omega $ (the unknown numerical coefficient
is the combination of the two factors: explicit dependence of the
result on the wavefunction $\psi _{b}$ and the unknown factor in
Eq.(\ref{eq:psidef})). 

To prove the equivalence of the Hamiltonians (\ref{eq:Hspinboson})
and (\ref{eq:Hbnew}) we follow the unitary transformation: $H^{\prime }=S^{-1}(H_{LL}+H_{b})S$
with $S=\exp (\sigma _{z}i\phi(0)/2)$ and identify $u=v_{s}$,
$\Delta =\tilde{\Omega }(\rho_al_{b})^{1/2}$, and 
\begin{equation}
\lambda _{q}=u\left|\frac{\pi q}{L}\right|^{1/2}
\left(\frac{g_{ab}}{2\pi u}(2K)^{1/2}-
\frac{1}{(2K)^{1/2}}\right).
\label{eq:lamgas}
\end{equation}
We note though, that the equivalence of the initial spin-boson Hamiltonian
(\ref{eq:Hspinboson}) and its quantum gases version (\ref{eq:Hbnew}),
(\ref{eq:HLL0}) can only be justified if for $ql_{b}\ll 1$. For
larger values of $q$ the interaction becomes more sophisticated,
but also decreases as soon as $ql_{b}\agt 1$. One can still use Eq.(\ref{eq:Hbnew}),
assuming that the summation over $q$ is restricted by the condition
$\omega (q)\alt \omega _{c}$, where the quantity $\omega _{c}\sim u/l_{b}$
plays the role of the high-frequency cutoff.

\begin{figure}
\centerline{\epsfig{file=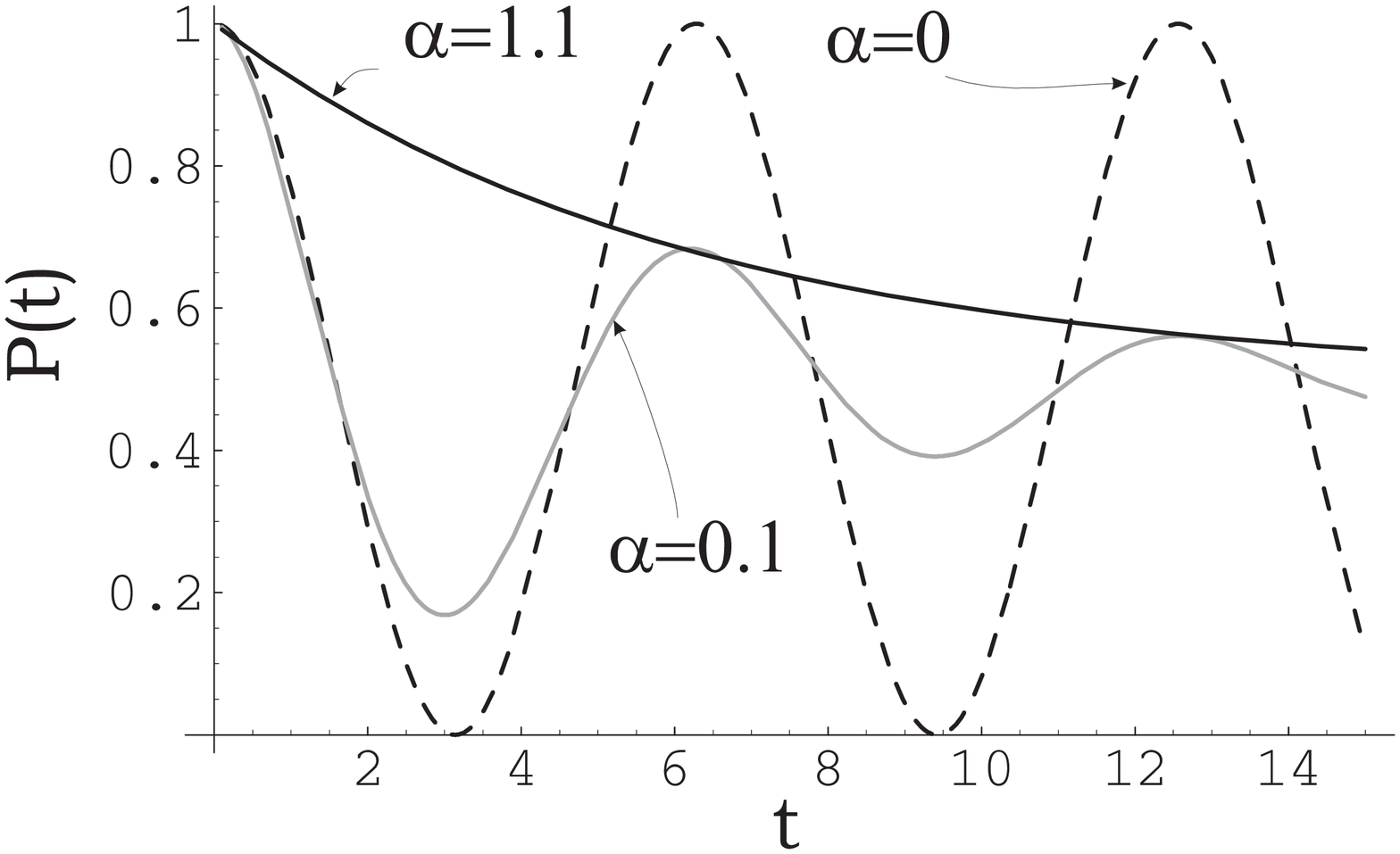,width=7.5cm}}
\caption{The atoms $b$ population relaxation. In the coherent
regime, at zero temperature, for $\alpha =0.1$ and $\alpha=0$. 
In the incoherent regime, at low temperature, for $\alpha=1.1$.}
\label{rabios}
\end{figure}

The spin-boson Hamiltonian (\ref{eq:Hspinboson})
can also describe an isolated spin-$1/2$ impurity dynamics in a 3d
Fermi system (the so called Kondo model). A slightly generalized (anisotropic)
version of the model can be formulated with the help of the following
Hamiltonian:
\begin{equation}
H=\sum _{q,\sigma }\epsilon (q)c_{\sigma q}^{\dagger }c_{\sigma q}+
J_{\Vert }S_{z}s_{z}(0)+J_{\perp }S_{\perp }s_{\perp }(0),
\label{eq:Kondoorig}
\end{equation}
 where $c_{\sigma q}$ is the annihilation of a fermion with (now
3d) momentum $q$ and the spin projection $\sigma $, $\mathbf{S}$
is the spin of the impurity, $\mathbf{s}$ is the total spin of the
electrons at a given point. The exchange couplings $J_{\perp }$and
$J_{\Vert }$ describe the interaction of the different spin components
($S_{\perp }s_{\perp }=S_{x}s_{x}+S_{y}s_{y}$). The exact solution of
the model is given in \cite{tsvelik}. Mapping of the Hamiltonian
(\ref{eq:Kondoorig}) to the spin-boson model Hamiltonian (\ref{eq:Hspinboson})
is described in every detail in \cite{spinboson:leggett,spinboson:costi}, and is based
on the fact that the spin density excitations of a free Fermi gas can serve
as an oscillator bath. This  leads to the identifications
$\Delta \sim J_{\perp }$, $u=v_{F}$ with $v_{F}$ being the Fermi
velocity of the Fermi gas, and 

\begin{equation}
\lambda _{q}=u\left(\frac{\pi q}{L}\right)^{1/2}\left(1-\frac{J_{\Vert }}{4\pi u}
\right).
\label{eq:lamKondo}
\end{equation}
The behavior of the system (the character of the impurity spin dynamics)
depends strongly on the sign of the coupling $J_{\Vert }$: the situations
with $J_{\Vert }<0$ and $J_{\Vert }>0$ are commonly referred to
as the ferromagnetic and anti-ferromagnetic cases respectively.

The analysis above shows, that the spin-boson model (\ref{eq:Hspinboson})
can describe both the Kondo and our quasi-1d Bose gas with an impurity.
On the other hand, this also means that the cold atoms version of
the spin-boson model can, among other things, be used to model the
Kondo Hamiltonian and the related phenomena, provided that the values
of the independent parameters $g_{ab}$ and $K$ in (\ref{eq:lamgas})
are selected accordingly, i.e. the couplings (\ref{eq:lamgas}) and
(\ref{eq:lamKondo}) match each other. Both systems are 
characterized by 
$
J(\omega)=2\alpha \omega 
$
in the $\omega\rightarrow 0$ limit, where 
\begin{equation}
\alpha =\frac{1}{2K}\left(\frac{g_{ab}K^2}{\pi ^{2}\rho _{a}}-1\right)^{2}.
\end{equation}
In the Kondo language, $\alpha >1$ and $\alpha <1$ correspond to
the antiferromagnetic and ferromagnetic cases, respectively. We note,
that since both $K$ and $g_{ab}$ are independent and practically
unrestricted ($g_{ab}$ can have arbitrary sign and practically any
value, whereas $1<K<\infty $), the parameter $\alpha $ can take
arbitrary values, both small and large. For realistic values $g_{ab}\ll\rho_a$
$\alpha\approx 1/2K$ and is small. For example, for a potassium condensate
of $N_a=10^4$ particles the interaction parameter $g_{ab}/n_a \sim 10^{-2}$, 
and $K\sim 30$, which means that $\alpha \sim 0.02$ and can be made larger by
a Feshbach resonance. 

The dynamics of the ``spin impurities'' in our proposal
can be observed by following the population dynamics of the atoms
$b$ in the presence of the laser light $\Omega $. In the absence
of interaction between the ``impurity'' $b$ and the LL of atoms $a$, 
the atom of impurity should
undergo familiar undamped Rabi oscillations with the laser field. 
The actual damping is
determined by the width of the involved atomic levels and will be
neglected here. On the other hand, the interaction between the atoms
$a$ collectivize the Rabi oscillations of the ``spin'' and the
collective degrees of the LL of atoms $a$.

A particular interesting regime of the ohmic two-state system is the low $T$
and $\alpha<1/2$ weak damping case. In this limit the occupation $P(t)$ of, 
say, the state "$b$" exhibits damped Rabi oscillations \cite{diss-qft}
$$
P(t)=\cos\left(\Delta_r t \cos(\eta)\right)
\exp(-\Delta_r t \sin(\eta)),
$$
where $\eta=\pi\alpha/(2(1-\alpha))$, $\Delta _{r}=\Delta (\Delta /\omega _{c})^{\alpha /(\alpha -1)}$
(see Fig. \ref{rabios}). This result holds as long as $\alpha T\alt \Delta_r$.
At higher $T$ the dynamics is incoherent (no oscillations should be visible).

For $1/2<\alpha<1$ the system shows no oscillations and $P(t)$ is a sum of 
exponentials \cite{diss-qft}. This is the regime which is directly relevant 
for the Kondo model. The detailed description of the model dynamics is 
still unsolved.

The situation becomes simpler again for $\alpha>1$, which is the ferromagnetic 
analogue of the Kondo problem. At $T=0$ the spin is localized, whereas at larger
$T$ there are incoherent transitions between the states at a rate 
$\sim T^{2\alpha-1}$. This peculiar behavior was in fact observed in SQUIDs 
experiments \cite{squid3} (see the Fig.\ref{rabios}).

In fact the suggested experimental scheme is not confined to a case of a single 
spin impurity in a quantum liquid. If the $N_b$ is sufficiently large, the pseudo-spins 
can interact with "host" liquid in a collective manner. This situation is described
by Anderson model and is characteristic
to a number of important solid state systems,
like heavy fermionic compounds (see e.g. \cite{tsvelik}). 
The spin-boson model is also a paradigmatic
model for qubits damped by an environment. Understanding
how such a system reacts under {\it tunable} conditions,
is obviously important for all those interested
in quantum computation.

Discussions with U. R. Fischer, J.I. Cirac, M. Cazalilla, D. Jaksch
are gratefully acknowledged. Work at Innsbruck
supported in part by the A.S.F., EU
Networks, and the Institute for Quantum Information. P.Z. thanks 
the Alexander von Humboldt Award for support.
A.R. has been supported by the 
European Commission Project Cold Quantum Gases RTN
Network Contract No. HPRN-CT-2000-00125.

\end{document}